\documentstyle[12pt]{article}
\begin{document}
\title{Involutive orbits of non-Noether symmetry groups}
\author{George Chavchanidze} \date{} \maketitle
\thanks{Department of Theoretical Physics, A. Razmadze Institute of Mathematics,
1 Aleksidze Street, Tbilisi 0193, Georgia}
\begin{abstract}{\bf Abstract.} We consider set of functions on Poisson manifold
related by continues one-parameter group of transformations. Class
of vector fields that produce involutive families of functions is
investigated and relationship between these vector fields and
non-Noether symmetries of Hamiltonian dynamical systems is
outlined. Theory is illustrated with sample models: modified
Boussinesq system and Broer-Kaup system.\end{abstract} {\bf
Keywords:} Non-Noether symmetry; Conservation laws; Modified
Boussinesq
system; Broer-Kaup system;
\\
In Hamiltonian integrable models, conservation laws often form
involutive orbit of one-parameter symmetry group. Such a symmetry
carries important information about integrable model and its
bi-Hamiltonian structure. The present paper is an attempt to
describe class of one-parameter group of transformations of
Poisson manifold that possess involutive orbits and may be related
to Hamiltonian
integrable systems. \\
Let $C^{\infty }(M)$ be algebra of smooth functions on manifold
$M$ equipped with Poisson bracket
\begin{eqnarray}
\label{eq:e1} \{f , g\} = W(df \wedge dg)
\end{eqnarray}
where $W$ is Poisson bivector satisfying property $[W , W] = 0$.
Each vector field $E$ on manifold $M$ gives rise to one-parameter
group of transformations of $C^{\infty }(M)$ algebra
\begin{eqnarray}
\label{eq:e2} g_{z} = e^{zL_{E}}
\end{eqnarray}
where $L_{E}$ denotes Lie derivative along the vector field $E$.
To any smooth function $J \in C^{\infty }(M)$ this group assigns
orbit that goes through $J$
\begin{eqnarray}
\label{eq:e3} J(z) = g_{z}(J) = e^{zL_{E}}(J) = J + zL_{E}J +
\frac{1}{2}z^2 (L_{E})^2 J + \cdots
\end{eqnarray}
the orbit $J(z)$ is called involutive if
\begin{eqnarray}
\label{eq:e4} \{J(x) , J(y)\} = 0~~~~~~~~~~\forall x, y \in R
\end{eqnarray}
Involutive orbits are often related to integrable models where
$J(z)$ plays the
role of involutive family of conservation laws. \\
Involutivity of orbit $J(z)$ depends on nature of vector field $E$
and function $J = J(0)$ and in general it is hard to describe all
pairs $(E , J)$ that produce involutive orbits however one
interesting class of involutive orbits can
be outlined by the following theorem: \\
{\bf Theorem 1.} For any non-Poisson $[E , W] \neq 0$ vector field
$E$ satisfying property
\begin{eqnarray}
\label{eq:e5} [E , [E , W]] = 0
\end{eqnarray}
and any function $J$ such that
\begin{eqnarray}
\label{eq:e6} W(dL_{E}J) = c[E , W](dJ)~~~~~~~~~~c \in R\setminus
(0\cup N)
\end{eqnarray}
one-parameter family of functions $J(z) = e^{zL_{E}}(J)$ is involutive. \\
{\bf Proof.} By taking Lie derivative of property (6) along the
vector field $E$ we get
\begin{eqnarray}
\label{eq:e7} [E , W](dL_{E}J) + W(d(L_{E})^2 J) = c[E,[E ,
W]](dJ) + c[E , W](dL_{E}J)
\end{eqnarray}
where $c$ is real constant which is neither zero nor positive
integer. Taking into account (5) one can rewrite result as follows
\begin{eqnarray}
\label{eq:e8} W(d(L_{E})^2 J) = (c - 1)[E , W](dL_{E}J)
\end{eqnarray}
that after $m$ iterations produces
\begin{eqnarray}
\label{eq:e9} W(d(L_{E})^{m + 1}J) = (c - m)[E , W](d(L_{E})^{m}J)
\end{eqnarray}
Now using this property let us prove that functions $J^{(m)} =
(L_{E})^{m}J$ are in involution. Indeed
\begin{eqnarray}
\label{eq:e10} \{J^{(k)}, J^{(m)}\} = W(dJ^{(k)} \wedge dJ^{(m)})
\end{eqnarray}
Suppose that $k > m$ and let us rewrite Poisson bracket as follows
\begin{eqnarray}
\label{eq:e11}
&& W(dJ^{(k)} \wedge dJ^{(m)}) = W(d(L_{E})^{k}J \wedge dJ^{(m)}) = L_{W(d(L_{E})^{k}J)}J^{(m)} \nonumber \\
&& = (c - k + 1)L_{[E , W](d(L_{E})^{k - 1}J)}J^{(m)} = (c - k +
1)[E , W](dJ^{(k
- 1)} \wedge dJ^{(m)}) \nonumber \\
&& = - (c - k + 1)L_{[E , W](d(L_{E})^{m}J)}J^{(k - 1)} = - {c - k
+ 1\over c -
m}L_{W(d(L_{E})^{m + 1}J)}J^{(k - 1)} \nonumber \\
&& = {c - k + 1\over c - m}W(dJ^{(k - 1)} \wedge dJ^{(m + 1)})
\end{eqnarray}
Thus we have
\begin{eqnarray}
\label{eq:e12} (c - m)\{J^{(k)}, J^{(m)}\} = (c - k + 1)\{J^{(k -
1)}, J^{(m + 1)}\}
\end{eqnarray}
Using this property $2(m - k)$ times produces
\begin{eqnarray}
\label{eq:e13} \{J^{(k)}, J^{(m)}\} = \{J^{(m)}, J^{(k)}\}
\end{eqnarray}
and since Poisson bracket is skew-symmetric we finally get
\begin{eqnarray}
\label{eq:e14} \{J^{(k)}, J^{(m)}\} = 0
\end{eqnarray}
So we showed that functions $J^{(m)} = (L_{E})^{m}J$ are in
involution. In the same time orbit $J(z)$ is linear combination of
functions $J^{(m)}$ and thus it
is involutive as well. \\
{\bf Remark.} Property (9) implies that vector field
\begin{eqnarray}
\label{eq:e15} S = (c - m)E + t(c - m + 1)W(dJ^{(m + 1)})
\end{eqnarray}
is non-Noether symmetry [1] of Hamiltonian dynamical system
\begin{eqnarray}
\label{eq:e16} {d\over dt}f = \{J^{(m)}, f\}
\end{eqnarray}
in other words non-Poisson vector field $S$ commutes with time
evolution defined by Hamiltonian vector field
\begin{eqnarray}
\label{eq:e17} X = {\partial \over \partial t} + W(dJ^{(m)})
\end{eqnarray}
This fact can be checked directly
\begin{eqnarray}
\label{eq:e18}
&& [S , X] = (c - m)[E , X] + t(c - m + 1)[W(dJ^{(m + 1)}), W(dJ^{(m)})] \nonumber \\
&& - (c - m + 1)W(dJ^{(m + 1)}) = (c - m)[E , W](dJ^{(m)}) + (c - m)W(dL_{E}J^{(m)}) \nonumber \\
&& + t(c - m + 1)W(d\{J^{(m + 1)},J^{(m)}\}) - (c - m + 1)W(dJ^{(m + 1)}) \nonumber \\
&& = W(dJ^{(m + 1)}) + (c - m)W(dJ^{(m + 1)}) - (c - m +
1)W(dJ^{(m + 1)}) = 0
\end{eqnarray}
In the same time property (9) means that functions $J^{(m)} =
(L_{E})^{m}J$ form Lenard scheme with respect to bi-Hamiltonian
structure formed by Poisson
bivector fields $W$ and $[E , W]$ (see [1],[4]). \\
In many infinite dimensional integrable Hamiltonian systems
Poisson bivector has nontrivial kernel, and set of conservation
laws belongs to orbit of non-Noether symmetry group that goes
through centre of Poisson algebra. This fact is
reflected in the following theorem: \\
{\bf Theorem 2.} If non-Poisson vector field $E$ satisfies
property
\begin{eqnarray}
\label{eq:e19} [E, [E , W]] = 0
\end{eqnarray}
then every orbit derived from centre $I$ of Poisson algebra
$C^{\infty }(M)$ is
involutive. \\
{\bf Proof.} If function $J$ belongs to centre $J \in I$ of
Poisson algebra $C^{\infty }(M)$ then by definition $W(dJ) = 0$.
By taking Lie derivative of this condition along vector field $E$
one gets
\begin{eqnarray}
\label{eq:e20} W(dL_{E}J) = - [E , W](dJ)
\end{eqnarray}
that according to Theorem~1 ensures involutivity of $J(z)$ orbit. \\
{\bf Sample.} The theorems proved above may have interesting
applications in theory of infinite dimensional Hamiltonian models
where they provide simple way to construct involutive family of
conservation laws. One non-trivial example of such a model is
modified Boussinesq system [2],[5],[6] described by the following
set of partial differential equations
\begin{eqnarray}
\label{eq:e21}
&& u_{t} = cv_{xx} + u_{x}v + uv_{x} \nonumber \\
&& v_{t} = - cu_{xx} + uu_{x} + 3vv_{x}
\end{eqnarray}
where $u = u(x, t), v = v(x, t)$ are smooth functions on $R^2 $
subjected to zero boundary conditions $u(\pm \infty , t) = v(\pm
\infty , t) = 0$ This system can be rewritten in Hamiltonian form
\begin{eqnarray}
\label{eq:e22} {d\over dt}f = \{h, f\} = W(dh \wedge df)
\end{eqnarray}
with the following Hamiltonian
\begin{eqnarray}
\label{eq:e23} h = {1\over 2}\int _{- \infty }^{+ \infty } (u^2 v
+ v^3 + 2cuv_{x})dx
\end{eqnarray}
and Poisson bracket defined by Poisson bivector field
\begin{eqnarray}
\label{eq:e24} W = \int _{- \infty }^{+ \infty } \frac{1}{2}(A
\wedge A_{x} + B \wedge B_{x})dx
\end{eqnarray}
where $A, B$ are vector fields that for every smooth functional $R
= R(u, x)$ are defined via variational derivatives $A(R) = \delta
R/\delta u$ and $B(R) = \delta R/\delta v$. For Poisson bivector
(24) there exist vector field $E$ such that
\begin{eqnarray}
\label{eq:e25} [E,[E,W]] = 0
\end{eqnarray}
this vector field has the following form
\begin{eqnarray}
\label{eq:e26} && E = \int _{- \infty }^{+ \infty } (uvA_{x} -
cvA_{xx} + (uu_{x} + vv_{x})B +
(u^2 + 2v^2 )B_{x} + cuB_{xx})xdx \nonumber \\
&& = - \int _{- \infty }^{+ \infty } [(uv + 2cv_{x} + x((uv)_{x} + cv_{xx}))A \nonumber \\
&& + (u^2 + 2v^2 - 2cu_{x} + x(uu_{x} + 3vv_{x} - cu_{xx}))B]dx
\end{eqnarray}
Applying one-parameter group of transformations generated by this
vector field to centre of Poisson algebra which in our case is
formed by functional
\begin{eqnarray}
\label{eq:e27} J = \int _{- \infty }^{+ \infty } (ku + mv)dx
\end{eqnarray}
where $k, m$ are arbitrary constants, produces involutive orbit
that recovers infinite sequence of conservation laws of modified
Boussinesq hierarchy
\begin{eqnarray}
\label{eq:e28}
&& J^{(0)} = \int _{- \infty }^{+ \infty } (ku + mv)dx \nonumber \\
&& J^{(1)} = L_{E}J^{(0)} = {m\over 2}\int _{- \infty }^{+ \infty }(u^2 + v^2 )dx \nonumber \\
&& J^{(2)} = (L_{E})^2 J^{(0)} = m\int _{- \infty }^{+ \infty } (u^2 v + v^3 + 2cuv_{x})dx \nonumber \\
&& J^{(3)} = (L_{E})^3 J^{(0)} = {3m\over 4}\int _{- \infty }^{+
\infty } (u^{4}
+ 5v^{4} + 6u^2 v^2 \nonumber \\
&& - 12cv^2 u_{x} + 4c^2 u_{x}^2 + 4c^2 v_{x}^2 )dx \nonumber \\
&& J^{(m)} = (L_{E})^{m}J^{(0)} = L_{E}J^{(m - 1)}
\end{eqnarray}
\\
{\bf Sample.} Another interesting model that has infinite sequence
of conservation laws lying on single orbit of non-Noether symmetry
group is Broer-Kaup system [3],[5],[6], or more precisely special
case of Broer-Kaup system formed by the following partial
differential equations
\begin{eqnarray}
\label{eq:e29}
&& u_{t} = cu_{xx} + 2uu_{x} \nonumber \\
&& v_{t} = - cv_{xx} + 2uv_{x} + 2u_{x}v
\end{eqnarray}
where $u = u(x, t), v = v(x, t)$ are again smooth functions on
$R^2 $ subjected to zero boundary conditions $u(\pm \infty , t) =
v(\pm \infty , t) = 0$ Equations (29) can be rewritten in
Hamiltonian form
\begin{eqnarray}
\label{eq:e30} {d\over dt}f = \{h, f\} = W(dh \wedge df)
\end{eqnarray}
with the Hamiltonian equal to
\begin{eqnarray}
\label{eq:e31} h = \int _{- \infty }^{+ \infty } (u^2 v +
cu_{x}v)dx
\end{eqnarray}
and Poisson bracket defined by
\begin{eqnarray}
\label{eq:e32} W = \int _{- \infty }^{+ \infty } A \wedge B_{x}dx
\end{eqnarray}
One can show that the following vector field $E$
\begin{eqnarray}
\label{eq:e33} && E = \int _{- \infty }^{+ \infty }(u^2 A_{x} -
cuA_{xx} + (uv)_{x}B + 3uvB_{x}
+ cvB_{xx})xdx \nonumber \\
&& = - \int _{- \infty }^{+ \infty } [(u^2 + 2cu_{x} + x(2uu_{x} + cu_{xx}))A \nonumber \\
&& + (3uv - 2cv_{x} + x(2(uv)_{x} - cv_{xx}))B]dx
\end{eqnarray}
has property
\begin{eqnarray}
\label{eq:e34} [E,[E,W]] = 0
\end{eqnarray}
and thus group of transformations generated by this vector field
transforms centre of Poisson algebra formed by functional
\begin{eqnarray}
\label{eq:e35} J = \int _{- \infty }^{+ \infty } (ku + mv)dx
\end{eqnarray}
into involutive orbit that reproduces well known infinite set of
conservation laws of modified Broer-Kaup hierarchy
\begin{eqnarray}
\label{eq:e36}
&& J^{(0)} =\int _{- \infty }^{+ \infty } (ku + mv)dx \nonumber \\
&& J^{(1)} = L_{E}J^{(0)} = m\int _{- \infty }^{+ \infty } uvdx \nonumber \\
&& J^{(2)} = (L_{E})^2 J^{(0)} = 2m\int _{- \infty }^{+ \infty } (u^2 v + cu_{x}v)dx \nonumber \\
&& J^{(3)} = (L_{E})^3 J^{(0)} = 3m\int _{- \infty }^{+ \infty }
(2u^3 v - 3cu^2
v_{x} - 2c^2 u_{x}v_{x})dx \nonumber \\
&& J^{(m)} = (L_{E})^{m}J^{(0)} = L_{E}J^{(m - 1)}
\end{eqnarray}
\\
Two samples discussed above are representatives of one interesting
family of infinite dimensional Hamiltonian systems formed by $D$
partial differential equations of the following type
\begin{eqnarray}
\label{eq:e37} && U_{t} = - 2FGU_{xx} + \langle U , GU_{x}\rangle
C + \langle C , GU_{x}\rangle
U + \langle C , GU\rangle U_{x} \nonumber \\
&& detG \neq 0,~~~~~~~~~~G^{T} = G,~~~~~~~F^{T} = - F \nonumber \\
&& F_{mn}C_{k} + F_{km}C_{n} + F_{nk}C_{m} = 0
\end{eqnarray}
where $U$ is vector with components $u_{m}$ that are smooth
functions on $R^2 $ subjected to zero boundary conditions
\begin{eqnarray}
\label{eq:e38} && u_{m} = u_{m}(x, t);~~~~~~~~~~u_{m}(\pm \infty ,
t) = 0;~~~~~~~~~~m = 1 ... D
\end{eqnarray}
$G$ is constant symmetric nondegenerate matrix, $F$ is constant
skew-symmetric matrix, $C$ is constants vector that satisfies
condition
\begin{eqnarray}
\label{eq:e39} F_{mn}C_{k} + F_{km}C_{n} + F_{nk}C_{m} = 0
\end{eqnarray}
and $\langle · , · \rangle $ denotes scalar product
\begin{eqnarray}
\label{eq:e40} \langle X , Y\rangle = \sum _{m=1}^{D}X_{m}Y_{m}.
\end{eqnarray}
System of equations (37) is Hamiltonian with respect to Poisson
bivector equal to
\begin{eqnarray}
\label{eq:e41} W = \int _{- \infty }^{+ \infty }\langle A ,
G^{-1}A_{x}\rangle dx
\end{eqnarray}
where $A$ is vector with components $A_{m}$ that are vector fields
defined for every smooth functional $R(u)$ via variational
derivatives $A_{m}(R) = \delta R/\delta u_{m}$. Moreover this
model is actually bi-Hamiltonian as there exist another invariant
Poisson bivector
\begin{eqnarray}
\label{eq:e42} \hat{W} = \int _{- \infty }^{+ \infty }\{\langle C
, A\rangle \langle U , A_{x}\rangle + \langle A_{x} ,
FA_{x}\rangle \}dx
\end{eqnarray}
that is compatible with $W$ or in other words
\begin{eqnarray}
\label{eq:e43} [W , W] = [W , \hat{W}] = [\hat{W} , \hat{W}] = 0
\end{eqnarray}
Corresponding Hamiltonians that produce Hamiltonian realization
\begin{eqnarray}
\label{eq:e44} {d\over dt}U = \hat{W}(d\hat{H} \wedge dU) = W(dH
\wedge dU)
\end{eqnarray}
of the evolution equations (37) are
\begin{eqnarray}
\label{eq:e45} \hat{H} = \frac{1}{2}\int _{- \infty }^{+ \infty
}\langle U , GU\rangle dx
\end{eqnarray}
and
\begin{eqnarray}
\label{eq:e46} H = {1\over 2}\int _{- \infty }^{+ \infty
}\{\langle C , GU\rangle \langle U , GU\rangle + 2\langle FGU_{x}
, GU\rangle \}dx
\end{eqnarray}
The most remarkable property of system (37) is that it possesses
set of conservation laws that belong to single orbit obtained from
centre of Poisson algebra via one-parameter group of
transformations generated by the following vector field
\begin{eqnarray}
\label{eq:e47} && E = \int _{- \infty }^{+ \infty }\{\langle C ,
GU\rangle \langle U , A_{x}\rangle
+ \langle U , GU\rangle \langle C , A_{x}\rangle \nonumber \\
&& + \langle U , GU_{x}\rangle \langle C , A\rangle + 2\langle FGU
, A_{xx}\rangle
\}xdx \nonumber \\
&& = \int _{- \infty }^{+ \infty }\{\langle C , GU\rangle \langle
U , A\rangle +
\langle U , GU\rangle \langle C , A\rangle + 4\langle FGU_{x} , A\rangle \nonumber \\
&& + x (\langle C , GU_{x}\rangle \langle U , A\rangle + \langle C
, GU\rangle \langle
U_{x} , A\rangle \nonumber \\
&& + \langle U , GU_{x}\rangle \langle C , A\rangle + 2\langle
FGU_{xx} , A\rangle )\}dx
\end{eqnarray}
Note that centre of Poisson algebra (with respect to bracket
defined by $W$) is formed by functionals of the following type
\begin{eqnarray}
\label{eq:e48} J = \int _{- \infty }^{+ \infty }\langle K ,
U\rangle dx
\end{eqnarray}
where $K$ is arbitrary constant vector and applying group of
transformations generated by $E$ to this functional $J$ yields the
infinite sequence of functionals
\begin{eqnarray}
\label{eq:e49}
&& J^{(0)} = \int _{- \infty }^{+ \infty }\langle K , U\rangle dx \nonumber \\
&& J^{(1)} = L_{E}J^{(0)} = \frac{1}{2}\langle C , K\rangle \int
_{- \infty }^{+
\infty }\langle U , GU\rangle dx \nonumber \\
&& J^{(2)} = (L_{E})^2 J^{(0)} = \langle C , K\rangle \int _{-
\infty }^{+ \infty
}\{\langle C , GU\rangle \langle U , GU\rangle + 2\langle FGU_{x} , GU\rangle \}dx \nonumber \\
&& J^{(3)} = (L_{E})^3 J^{(0)} = \frac{1}{4}\langle C , K\rangle
\int _{- \infty
}^{+ \infty }\{3\langle C , GC\rangle \langle U , GU\rangle ^2 \nonumber \\
&& + 12\langle C , GU\rangle ^2 \langle U , GU\rangle + 32\langle
C , GU\rangle
\langle GU , FGU_{x}\rangle \nonumber \\
&& + 24\langle U , GU\rangle \langle GC , FGU_{x}\rangle +
48\langle FGU_{x} ,
GFGU_{x}\rangle \}dx \nonumber \\
&& J^{(m)} = (L_{E})^{m}J^{(0)} = L_{E}J^{(m - 1)}
\end{eqnarray}
One can check that the vector field $E$ satisfies condition
\begin{eqnarray}
\label{eq:e50} [E , [E , W]] = 0
\end{eqnarray}
and according to Theorem~2 the sequence $J^{(m)}$ is involutive.
So $J^{(m)}$ are conservation laws of bi-Hamiltonian dynamical
system (37) and vector field $E$ is related to non-Noether
symmetries of evolutionary equations (see Remark 1).
\\
Note that in special case when $C, F, G, K$ have the following
form
\begin{eqnarray}
\label{eq:e51} D = 2,~~~~~~F_{12} = - F_{21} =
\frac{1}{2}c,~~~~~~C = K = (0 , 1),~~~~~~G = 1
\end{eqnarray}
model (37) reduces to modified Boussinesq system discussed above.
Another choice of constants $C, F, G, K$
\begin{eqnarray}
\label{eq:e52} D = 2,~~~~~F_{12} = - F_{21} = \frac{1}{2}c,~~~~~~C
= K = (0 , 1)\nonumber \\
G_{12} = G_{21} = 1 ,~~~~~G_{11} = G_{22} = 0
\end{eqnarray}
gives rise to Broer-Kaup system described in previous sample. \\
{\bf Conclusions.} Groups of transformations of Poisson manifold
that possess involutive orbits play important role in some
integrable models where conservation laws form orbit of
non-Noether symmetry group. Therefore classification of vector
fields that generate such a groups would create good background
for description of remarkable class of integrable system that have
interesting geometric origin. The present paper is an attempt to
outline one particular class of vector fields that are related to
non-Noether symmetries of Hamiltonian dynamical systems and
produce involutive families of conservation
laws. \\
{\bf Acknowledgements.} The research described in this publication
was made possible in part by Award No. GEP1-3327-TB-03 of the
Georgian Research \& Development Foundation (GRDF) and the U.S.
Civilian Research \& Development
Foundation for the Independent States of the Former Soviet Union (CRDF). \\

\end{document}